\documentstyle[sprocl_pre,a4,12pt,epsfig]{article}

\newenvironment{licompact}
  {\begin{list}{$\bullet$}
    {\setlength{\parsep}{0.4ex}
     \setlength{\topsep}{0.6ex}
     \setlength{\itemsep}{0ex}
     \setlength{\leftmargin}{1.5em}}
  }
  {\end{list}}
\begin{document}

\title{DETERMINATION OF CRITICAL EXPONENTS \\ IN NUCLEAR SYSTEMS
\footnote{\normalsize To appear in the proceedings of the 1st Catania
    Relativistic Ion Studies: Critical Phenomena and Collective Observables,
    Acicastello, May 27-31, 1996.\\ \hspace*{1ex}}
}

\author{
 W.~F.~J.~M\"uller$^{1}$ for the ALADIN collaboration:
\\[0.5ex]
 R.~Bassini,$^{2}$
 M.~Begemann-Blaich,$^{1}$
 Th.~Blaich,$^{3}$
 H.~Emling,$^{1}$
 A.~Ferrero,$^{2}$
 S.~Fritz,$^{1}$
 C.~Gro\ss,$^{1}$
 G.~Imm\'{e},$^{4}$
 I.~Iori,$^{2}$
 U.~Kleinevo{\ss},$^{1}$
 G.~J.~Kunde,$^{1}$\footnote[2]{Present address:
  NSCL, Michigan State University,
  East Lansing, MI 48824}
 W.~D.~Kunze,$^{5}$
 V.~Lindenstruth,$^{1}$\footnote[3]{Present address:
  Nuclear Science Division, LBNL, Berkeley, CA 94720}
 U.~Lynen,$^{1}$
 M.~Mahi,$^{1}$
 A.~Moroni,$^{2}$
 T.~M\"ohlenkamp,$^{6}$
 B.~Ocker,$^{1}$
 T.~Odeh,$^{1}$
 J.~Pochodzalla,$^{1}$\footnote[4]{Present address:
  Max Planck Institut f\"ur Kernphysik, Heidelberg, Germany}
 G.~Raciti,$^{4}$
 Th.~Rubehn,$^{1\ddagger}$
 H.~Sann,$^{1}$
 M.~Schnittker,$^{1}$
 A.~Sch\"uttauf,$^{5}$
 C.~Schwarz,$^{1}$
 W.~Seidel,$^{6}$
 V.~Serfling,$^{1}$
 J.~Stroth,$^{1}$
 W.~Trautmann,$^{1}$
 A.~Trzcinski,$^{7}$
 G.~Verde,$^{4}$
 A.~W\"orner,$^{1}$
 E.~Zude,$^{1}$
 B.~Zwieglinski$^{7}$
}
\address{
\vspace*{1ex}
$^{1}$ Gesellschaft  f\"ur  Schwerionenforschung, D-64291 Darmstadt, Germany
\\
$^{2}$ Dipartimento di Fisica, Universit\`{a} di Milano and
    I.N.F.N., I-20133 Milano, Italy
\\
$^{3}$ Institut f\"ur Kernchemie, Universit\"at Mainz, D-55099 Mainz, Germany
\\
$^{4}$ Dipartimento di Fisica dell' Universit\`{a} and I.N.F.N.,
    I-95129 Catania, Italy
\\
$^{5}$ Institut f\"ur Kernphysik, Universit\"at Frankfurt,
    D-60486 Frankfurt, Germany
\\
$^{6}$ Forschungszentrum Rossendorf, D-01314 Dresden, Germany
\\
$^{7}$ Soltan Institute for Nuclear Studies, 00-681 Warsaw, Hoza 69, Poland
}
%
%
\maketitle
\abstracts{
Signatures of critical behaviour in nuclear fragmentation are often based on
arguments from percolation theory.
We demonstrate with general thermodynamic considerations and studies of the
Ising model that the reliance on percolation as a reference model bears the
risk of missing parts of the essential physics.
}
%
%
\section{Introduction}
\label{sec:intro}
The existence of phase transitions in nuclear matter was proposed more than 20
years ago in conjunction with the description of the structure of neutron
stars~\cite{Baym-75,Palmer-74}.
This led to the question whether phase separation~\cite{Jaqaman-83}, dynamical
instabilities~\cite{Bertsch-83} or critical phenomena~\cite{Bauer-85} play a
role in the disintegration of highly excited nuclei produced in proton or heavy
ion induced collisions.
The observation of a power law for the fragment size
distribution~\cite{Finn-82,Minich-82} was the first experimental hint for a
second order phase transition, but the determination of critical
exponents~\cite{Campi-86} or other parameters~\cite{Siemens-83} turned out to be
difficult with the mostly inclusive data available at that
time~\cite{Panagiotou-84,Porile-89,Campi-88}.
In the past 3 years, however, data from a new generation of experiments with an
almost complete coverage for the decay products of an excited system allowed for
rapid progress.
The observation of potential signals of a first order phase
transition~\cite{Pochodzalla-95,Moretto-96} and the attempt of a quantitative
determination of critical exponents~\cite{Gilkes-94} sparked an intense
discussion on what reliable signatures for a first or second order phase
transition in a small and dynamically evolving system are.

In the following we will take the recent claim~\cite{Gilkes-94,CRIS-Tincknell},
that a continuous phase transition with critical exponents consistent with those
of a liquid-gas system has been observed in the fragmentation of nuclei, at
face value and revisit the theoretical arguments used in this analysis.
%
%
\section{Critical Phenomena in a Nutshell}
\label{sec:nutshell}
Near a critical point, the order parameter $\phi$, associated susceptibility
$\chi$, specific heat $c$, and the correlation length $\xi$ exhibit a power law
dependence on two reduced control parameters $\epsilon$ and $h$:
\[
\begin{array}{r @{\;\;} c @{\;\;} l}
  \phi_{(\epsilon,h=0)} &=&    B \epsilon^{\beta}		\\[0.5ex]
  \phi_{(\epsilon=0,h)} &\sim& h^{1/\delta}			\\[0.5ex]
  \chi_{(\epsilon,h=0)} &=&    \Gamma \epsilon^{-\gamma}	\\[0.5ex]
  c_{(\epsilon,h=0)}    &=&    A \epsilon^{-\alpha}		\\[0.5ex]
  \xi_{(\epsilon,h=0)}  &=&    \xi_0 \epsilon^{-\nu}
\end{array}
\mbox{\hspace{2em} with \hspace{2em}}
\begin{array}{c @{\hspace{1em}} c @{\hspace{1em}} c @{\hspace{1em}} c}
  & \mbox{Liquid-Gas} & \mbox{Magnet} & \mbox{Percolation}	\\[0.5ex]
  \phi & \rho-\rho_c & M & P_\infty				\\[0.5ex]
  \chi & {\textstyle \frac{1}{\rho^2} \frac{\partial \rho}{\partial \mu}} &
	 {\textstyle \frac{\partial M}{\partial H}} &
	 {\textstyle \sum s^2 n_s}				\\[0.5ex]
  \epsilon & {\textstyle \frac{T-T_c}{T_c}} &
	 {\textstyle \frac{T-T_c}{T_c}} &
	 p_c - p						\\[0.5ex]
  h    & \mu - \mu_{0(T)} & H & p_g
\end{array}
\]
For liquid-gas, magnetic and percolation systems the order parameter $\phi$ is
the difference between liquid and critical density, magnetization per spin $M$
or the fraction of sites in the largest cluster $P_\infty$, respectively.
The susceptibility $\chi$ corresponds to the isothermal compressibility
$\kappa_T = -\frac{1}{V} \frac{\partial V}{\partial p}$ and the second moment of
the cluster size distribution for the liquid-gas and percolation case.
The `thermal' control parameter $\epsilon$ is the difference of the temperature
$T$ or bond probability $p$ to the critical value while the `field' control
parameter $h$ is given by the chemical potential $\mu$,
the external field $H$ or the ghost bond~\cite{Reynolds-77,Reynolds-80}
probability $p_g$.
%
%
\section{From Thermodynamics to Percolation and Back}
\label{sec:td2perc}
The exponents $\beta$ and $\gamma$ can be determined from a measurement
of $\phi_{(\epsilon,h=0)}$ and $\chi_{(\epsilon,h=0)}$.
For liquids it is indeed possible to determine $\Delta\rho = \rho - \rho_c$ and 
$\frac{\partial \rho}{\partial \mu}$ directly, e.g. by measuring the density
profile in the gravitational field~\cite{Wilcox-68,Moldover-79,Pestak-84}.

In the nuclear physics case, the only directly measured quantities are the
size and momenta of clusters produced in an interaction.
The standard argument to deduce $\Delta\rho$ and $\kappa_T$ from the cluster
size distribution uses the Fisher droplet model~\cite{Fisher-67}. In this
Ansatz the grandcanonical partition function ${\cal Z}$ is expressed as a sum
over cluster yields $Y_{(A)}$ which depend on a surface energy~$s$ and
chemical potential difference between liquid and gas phase $\mu$:
\[
    \ln {\cal Z} = \sum Y_{(A)}
\mbox{\hspace{2em} with \hspace{2em}}
    Y_{(A)} = q_0 A^{-\tau} \exp \left ( {\textstyle - \frac{1}{T}}
        \left ( s A^\sigma + \mu A \right ) \right )
\]
From this follows immediately that $\kappa_T$ is proportional to the
second moment of the cluster size distribution
\begin{equation}
    \kappa_T = \frac{T}{V \rho^2} \frac{\partial^2}{\partial \mu^2} \ln {\cal Z}
        = \frac{1}{T V \rho^2} \sum A^2 Y_{(A)} .
\label{kappa_m2_fisher}
\end{equation}
It can also be shown~\cite{Fisher-67} that $\Delta\rho$ is proportional to the
fraction of constituents belonging to the largest cluster $P_\infty$.
This strong similarity to percolation, where $P_\infty$ and the second moment
also play the role of order parameter and susceptibility, has led to a
widespread use of percolation in the modeling and interpretation of nuclear
physics experiments.
In particular, percolation has served as reference model for the development
of methods to extract critical exponents~\cite{Elliott-94,Woerner-95}.

Even though there is a mathematical connection between percolation and the
thermodynamics of interacting systems (the $q \rightarrow 1$ limit of the
Potts model corresponds to bond percolation~\cite{Kasteleyn-69}) there are
also some significant differences:
\begin{licompact}
\item Percolation is usually discussed in terms of only one control
      parameter, the bond or site probability, leaving the impression that
      the `field' control parameter, corresponding to chemical potential
      or average density in interacting systems, is of minor importance.
\item Finite size effects depend strongly on the ensemble in
      interacting systems~\cite{Promberger-95,CRIS-Hueller,CRIS-Gross}, but
      there is no direct equivalent to the concept of an ensemble in
      percolation.
\item Percolation theory simply starts with the definition of what a
      cluster is.
      Thermal systems, on the other hand, are usually defined in terms of the
      interaction between their constituents and the appropriate definition of
      a cluster is, as will be shown, nontrivial.
\end{licompact}
The almost exclusive reliance on percolation as a reference model bears
therefore the risk that important parts of the physics are missed.
In the following we will consequently turn back to the thermodynamic basics
and use the Ising model as guidance.
%
%
\section{From Constituents to Clusters}
\label{sec:con2clu}
As a first step it is interesting  to establish the connection between
$\kappa_T$ and the cluster size distribution with minimal assumptions and
without using a specific model.
We follow an idea given by Alexandrowicz~\cite{Alexandrowicz-89} for the
Ising model but generalize it to an arbitrary system.
In the grand canonical ensemble, $\kappa_T$ is related to the fluctuation of
the particle number $\sigma^2_N$ by
\[
  \sigma^2_N = \left . kT \frac{\partial N}{\partial \mu} \right | _{T,V}
             = kT \frac{N^2}{V} \kappa_T
\]
In the following, we consider a system with $N_S$~constituents coupled to a
reservoir with $N_R$~constituents:
\\[1ex]
\begin{center}
  \begin{picture}(300,50)(0,0)
    \put(100,10){$N_R$}
    \put(130,25){\oval(100,50)}
    \put(140,15){\framebox(30,20){$N_S$}}
    \put(0,35){$N_S$: System}
    \put(0,25){$N_R$: Reservoir}
    \put(0,15){$N = N_R + N_S$}
    \put(200,40){Define for constituent $i$:}
    \put(200,15){$  q_i = \left \{ \begin{array}{ll}
                    1 & \mbox{if in $S$} \\
                    0 & \mbox{if in $R$}
                  \end{array}    \right .
                $}
  \end{picture}
\end{center}
\vspace*{1ex}
To express $\sigma^2_{N_S}$ in terms of clusters we assume that the
constituents are grouped into clusters with the properties:
\\[0.5ex]
\hspace*{2em}{\bf P1} \hspace{1em}
    A cluster is either completely in $S$ or in $R$.
\\
\hspace*{2em}{\bf P2} \hspace{1em}
    Constituents in different clusters are uncorrelated.
\\[0.5ex]
$\sigma^2_{N_S}$ can now be rewritten as a $q$ correlation. The sum over
constituent pairs can be split into two parts, one where the pair is in the
same cluster and one where it is in different clusters:
\begin{eqnarray*}
  \sigma^2_{N_S} 
    & = & 
	  \left < N_S^2 \right > - \left < N_S \right >^2
\\
    & = & 
	  \left < \sum_{ij} (q_i-\overline{q})(q_j-\overline{q}) \right >
          \mbox{\hspace{3em} with
		$\overline{q} = \frac{\left < N_S \right >}{N}$
	       }
\\
    & = & 
	  \left < \sum_{\stackrel{ij}{\mbox{\scriptsize same}}}
          (q_i-\overline{q})(q_j-\overline{q}) \right > +
          \underbrace{
             \left < \sum_{\stackrel{ij}{\mbox{\scriptsize diff}}}
             (q_i-\overline{q})(q_j-\overline{q}) \right >
          }_{\stackrel{\mbox{\tiny (P2)}}{=} \; 0}
\\
    & \stackrel{\mbox{\tiny (P1)}}{=} & 
	 (1-\overline{q})^2
         \left < \sum_{\stackrel{c_i}{\mbox{\scriptsize S}}} |c_i|^2 \right > +
         \overline{q}^2
         \left < \sum_{\stackrel{c_i}{\mbox{\scriptsize R}}} |c_i|^2 \right >
\end{eqnarray*}
In the last line, the sum over constituents was rewritten as a sum over
clusters  ${\displaystyle
	    \sum_{\stackrel{ij}{\mbox{\scriptsize same}}} \rightarrow
            \sum_{c_i} |c_i|^2 }$
and split into two parts, running over $S$ and $R$, respectively.
In the thermodynamic limit, $N_R \rightarrow \infty$ , $\overline{q}
\rightarrow 0$, the second term goes to zero if the largest cluster in $R$
grows slower than $N_R$ (in other words: No condensation in $R$)
and one finally gets
\begin{equation}
  \kappa_T = \frac{V}{k T}\frac{\sigma^2_{N_S}}{\left < N_S \right >^2}
           = \frac{1}{k T \rho} \sum_s s^2 n_s
\label{kappa_cluster}
\end{equation}
where $n_s$ is the concentration of clusters of size $s$.
This is equivalent to Eqn.~(\ref{kappa_m2_fisher}) but was derived with only
one essential assumption, that constituents in different clusters are
uncorrelated.
%
%
\section{The Ising Model}
\label{sec:ising}
One of the simplest thermodynamic models with a phase transition is the Ising
model, given by the Hamiltonian:
\begin{equation}
    {\cal H} = -J \sum_{<i,j>} \sigma_i \sigma_j - H \sum_i \sigma_i
\label{ising_hamiltonian}
\end{equation}
It can be interpreted as a model for a magnet, with $\sigma = \pm 1$
representing up and down spins, or as a lattice model of a gas~\cite{Yang-52}
where $\sigma = \pm 1$ now indicates whether a site is occupied or empty.
The number of sites $N$, magnetization $M$ and external field $H$ correspond
in the lattice gas interpretation to the volume $V$, density $\rho$ and
chemical potential $\mu$, respectively.

Order parameter $\left<|M|\right>$ and susceptibilities $\chi$ and $\chi^\prime$
for $T \ge T_c$ and $T < T_c$, respectively, are defined in terms of
constituents as:
\begin{equation}
\begin{array}{rcl}
 \left<|M|\right> & = & {\textstyle \left < \left |
                        \frac{1}{N} \sum_i \sigma_i \right | \right > }
 \\[1.0ex]
 \chi & = & { \textstyle \frac{N}{kT} \left < M^2 \right > =
        \frac{N}{kT} \left (\frac{1}{N^2}
            \left < \sum_{ij} \sigma_i \sigma_j \right > \right ) }
 \\[1.0ex]
 \chi^\prime & = & { \textstyle \frac{N}{kT} \left (
    \left < M^2 \right > - \left < \left | M \right | \right > ^2 \right ) }
\end{array}                                               
\label{mchi_true}
\end{equation}
$M$ and $\chi$ can be expressed in terms of clusters using arguments
which are for $\chi$ analogous to the ones in the previous section.
A detailed derivation along those lines was given by de Meo~\cite{DeMeo-90}
while other authors arrived at the same result from different
viewpoints~\cite{Alexandrowicz-89,Wang-89,Hu-84,Roussenq-82}.
One obtains for the cluster observables $\widetilde{M}$, $\widetilde{\chi}$
and $\widetilde{\chi^\prime}$:
\begin{equation}
\begin{array}{cclcc}
  \widetilde{M} & := &
        \left < P_\infty \right > & \leq &
        \left < \left | M \right | \right >
\\[0.7ex]
  \widetilde{\chi} & := &
        \frac{1}{kT} \: \chi_p & = &
        \chi
\\[0.7ex]
  \widetilde{\chi^\prime} & := &
        \frac{1}{kT} \left ( \chi_p^\prime +
        N (\Delta P)^2 \right ) & \geq &
        \chi^\prime
\end{array}
\label{mchi_cluster}
\end{equation}
with the fluctuation of largest cluster $(\Delta P)^2 = \left < P^2_\infty
\right > - \left < P_\infty \right >^2$ and the percolation susceptibility
$\chi_p = {\textstyle \sum_m m^2 <n_m>}$ where $n_m$ is the number of clusters
per site with magnetization $m$.

While the order parameter is indeed given by the relative size of the
largest cluster, we observe two significant differences between
$\widetilde{\chi}$ and $\chi_p$:
\begin{licompact}
\item The trivial $\frac{1}{kT}$ factor (see also
      Eqn.~(\ref{kappa_m2_fisher}) and (\ref{kappa_cluster})) makes sure that
      the paramagnetic or ideal gas limit $\chi \propto \frac{1}{kT}$ is
      approached for $T \rightarrow \infty$ where $\chi_p \rightarrow 1$.
      Even though it does not affect the asymptotic power law behaviour
      for $T \rightarrow T_c$ it will change the effective exponents if a wide
      temperature range is considered, like in the nuclear exponent
      analysis~\cite{Gilkes-94}.
\item The susceptibitity $\widetilde{\chi^\prime}$ for $T < T_c$ is given
      by the sum of $\chi_p^\prime$, the second moment taken without the
      largest cluster, and the fluctuation of the largest cluster $(\Delta P)^2$.
\end{licompact}

\begin{figure}[tb]
  \centerline{
    \epsfig{file=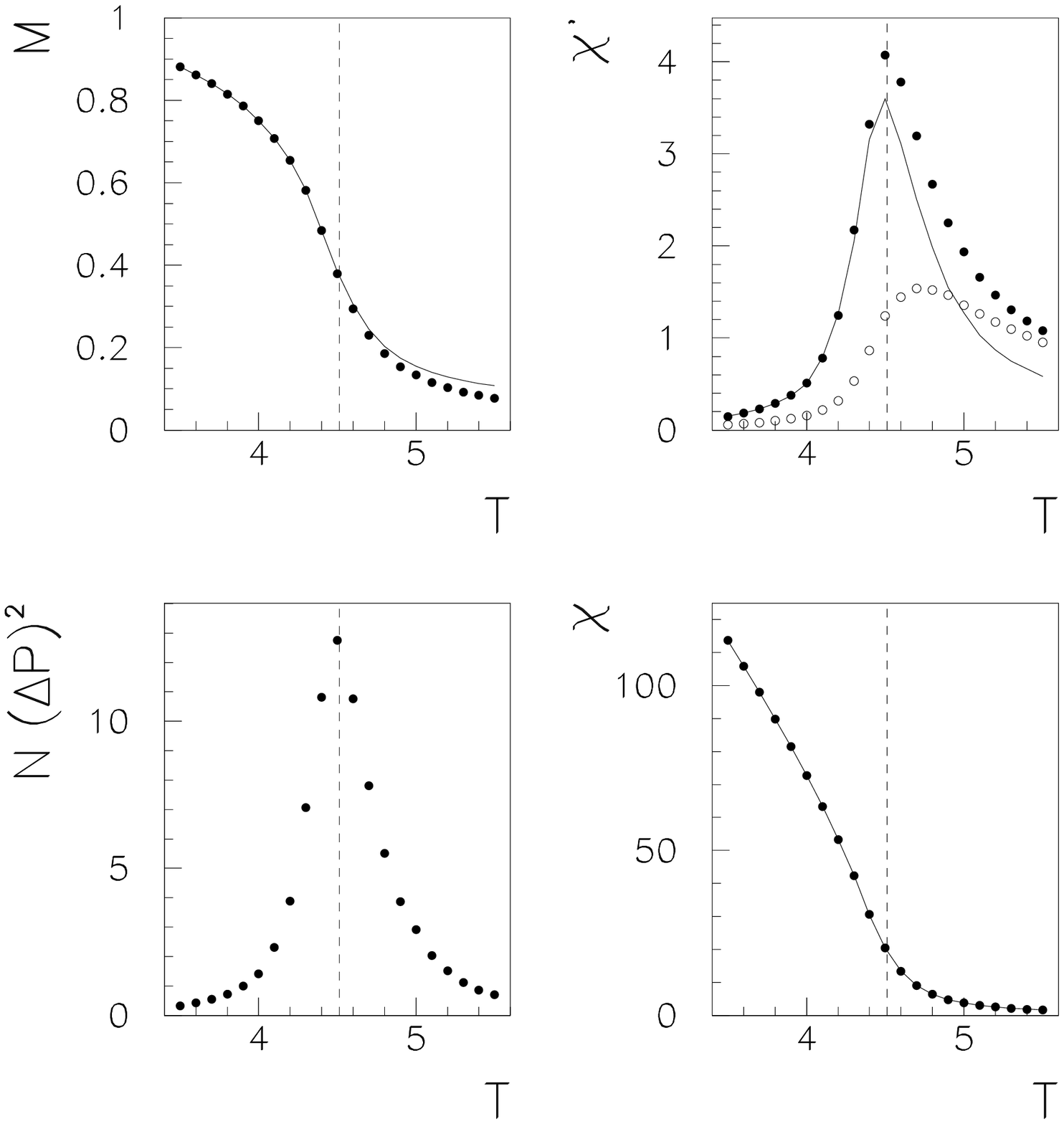,height=9.5cm}
    \epsfig{file=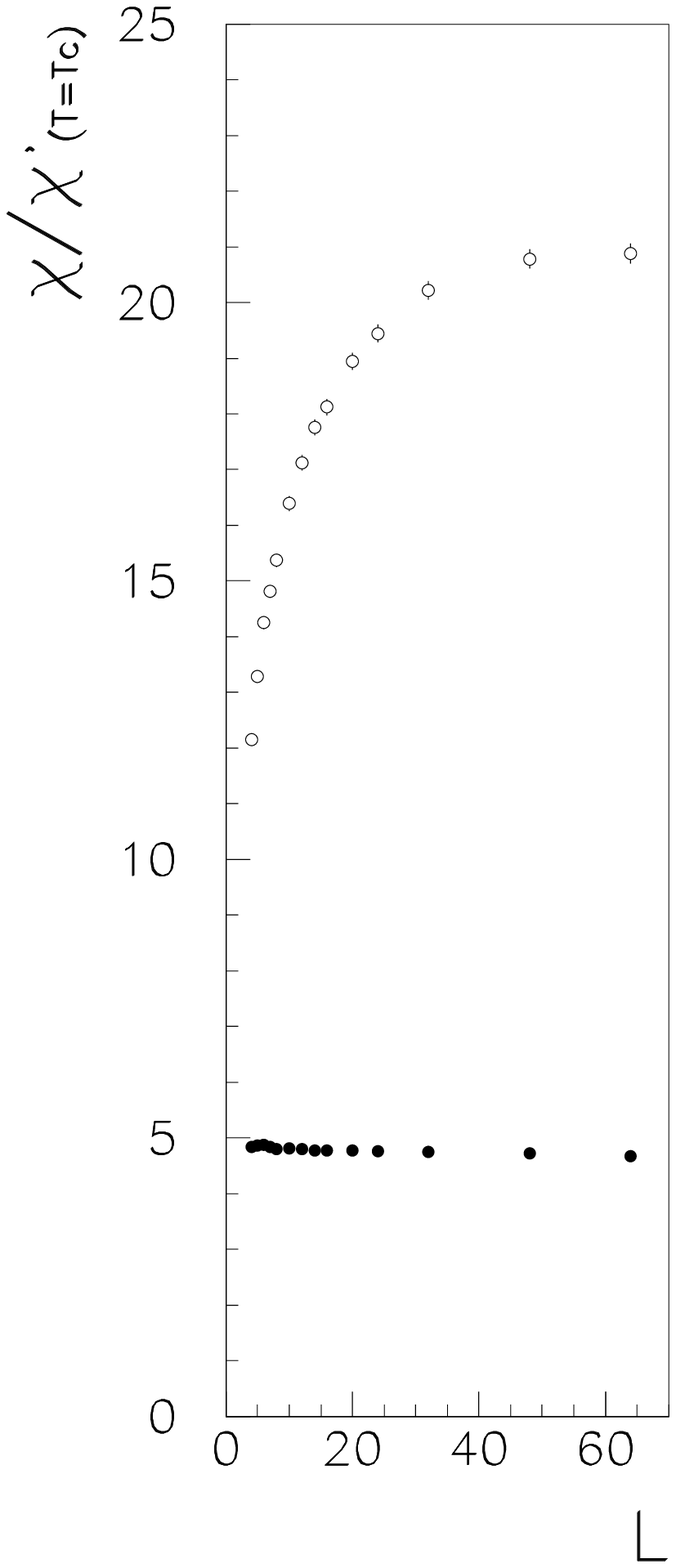,height=9.5cm}
  }
  \caption[]{
  Left two columns: Comparison of cluster observables (Eqn.~(\ref{mchi_cluster}),
  full dots) and true values (Eqn.~(\ref{mchi_true}), full lines) for an
  Ising model with $8^3$ sites. The dotted lines indicate the bulk critical
  temperature $T_c = 4.51152$~\cite{Bloete-95}.
\\
  Right column: Susceptibility ratio
  $\widetilde{\chi}/\widetilde{\chi^\prime}$ at $T_c$ as a function
  of the linear system size $L$ with (full dots) and without (open dots)
  the $(\Delta P)^2$ contribution to $\widetilde{\chi^\prime}$.
  }
  \label{fig_ising}
\end{figure}

\noindent
Numerical results for the 3d Ising model with periodic boundary conditions
are shown in Fig.~\ref{fig_ising}.
The calculations were done with the Swendsen-Wang~\cite{Swendsen-87} and
Wolff~\cite{Wolff-89} algorithms in implementations similar to
Ref.~\cite{Wang-90} using a random number generator proposed by
Ziff~\cite{Ziff-92,RNG-remark}.
In the left two columns we compare the cluster observables to the true values
given by Eqn.~(\ref{mchi_true}).
Consistent with Eqn.~(\ref{mchi_cluster}) we find $\widetilde{M}$ slightly
smaller than $\left < \left | M \right | \right >$ 
and $\widetilde{\chi}$ equal to $\chi$ while $\widetilde{\chi^\prime}$ is
only close to $\chi^\prime$ for $T < T_c$.

It is interesting to note, that the main contribution to
$\widetilde{\chi^\prime}$ is the fluctuation term $(\Delta P)^2$ while
the second moment term $\frac{1}{kT}\chi_p^\prime$ (indicated by open
dots in the top middle frame of Fig.~\ref{fig_ising}) carries only a small
part of the signal.
This can be seen more quantitatively in the right panel of Fig.~\ref{fig_ising}
where the susceptibility ratio $\widetilde{\chi}/\widetilde{\chi^\prime}$ at
$T_c$ is shown as a function of the linear system size $L$ with (full dots)
and without (open dots) the $(\Delta P)^2$ contribution to
$\widetilde{\chi^\prime}$.
Depending on system size, between 60 and 75\% of the susceptibility are
carried by the fluctuation term.
%
%
\section{Critical Clusters}
\label{sec:criclu}
Clusters can be defined in interacting systems in many different ways.
We call the result of a definition `critical clusters' if order parameter
and susceptibility correspond to size of largest cluster and second moment,
or formally, if Eqn.~(\ref{mchi_cluster}) holds.
We saw in the previous two sections that a necessary condition is the
noncorrelation of constituents in different clusters but have not given
so far an explicit definition.

The most obvious way to define a cluster is the `geometrical cluster' which
corresponds to a domain in the Ising model, or in the lattice gas picture
to a connected region of occupied sites surrounded by empty sites.
However, it is easy to see that this violates the noncorrelation
requirement~\cite{Alexandrowicz-89} and it has indeed been shown that
in 3d `geometrical clusters' lead to a percolation transition at temperatures
below the thermal transition~\cite{mueller-krum-74}, thus do not fulfill
Eqn.~(\ref{mchi_cluster}).

The proper prescription for `critical clusters' for the Ising model was given
by Coniglio and Klein~\cite{Coniglio-80} and requires that the geometrical
clusters are broken into smaller pieces with a bond percolation with 
$p_b = 1 - \exp(-\frac{2J}{kT})$.

The definition of a `physical cluster', based on pairwise binding, proposed
by Hill~\cite{Hill-55} and recently used in a lattice gas model for
nuclear fragmentation~\cite{Pan-95}, results in bond probabilities quite
similar to the Coniglio-Klein values.
Even though both definitions give numerically similar results for small
system sizes they are clearly not equivalent in the thermodynamic limit.
%
%
\section{Summary}
\label{sec:summary}
Percolation describes remarkably well the cluster distributions and
correlations in nuclear fragmentation~\cite{Desbois-87,Kreutz-93}.
This, together with its simplicity and ease of use, made it the ideal
reference model for the study of signatures of critical behaviour.
But given the hints, that we observe a liquid-gas rather than a percolation
phase transition, we have to face the limitations of percolation.
Even scratching at the surface of a proper thermodynamic description raises
many issues:
\begin{licompact}
\item {\bf Critical clusters} have rather remarkable properties, they are
      neither well separated nor compact \footnote{they are in general fractals 
      with a dimension $y_h = \frac{1}{\sigma\nu}$ or about 2.5 in the 3d Ising 
      class}, are interacting but have nevertheless no correlations between 
      constituents of different clusters.
      So one might wonder whether critical clusters are mere mathematical
      constructs, like the clusters in Mayer's cluster expansion, or real
      physical entities.
      All attempts to infer signatures of critical behaviour in nuclear physics
      not only imply that critical clusters are observable objects, they also
      assume that the distribution of clusters formed in the decay of a system
      is representative of the equilibrium distribution at some freeze-out
      condition.
      Even though there are attempts to support this connection with model
      calculations~\cite{Finocchiaro-96} it remains to be seen whether there is
      a more rigorous way to justify this.

\item {\bf Ensembles:}
      Although the experimental situation is certainly better represented
      by a microcanonical treatment we used for simplicity for all arguments
      in sections~\ref{sec:con2clu} and~\ref{sec:ising} the grand canonical
      ensemble.
      While the expectation values of extensive quantities do not depend on
      the ensemble in the thermodynamic limit it is easy to show that this is
      not the case for fluctuations~\cite{Moukarzel-89}.
      The cluster observables (Eqn.~(\ref{mchi_cluster})) are therefore likely
      to be ensemble dependent.
\item {\bf Role of control parameters:}
      The exponent analysis rests on the precondition that the temperature of
      the system is varied while the second control parameter, the density,
      stays constant and close to its critical value.
      Assuming that we observe conditions along a freeze-out line and taking
      Papp's schematic model~\cite{Papp-95,CRIS-Papp} as a guide one sees that
      the reverse might be true, that the temperature is almost constant but
      that the density varies.
      In this case one would determine different exponents, e.g.
      $1/\delta$ rather than $\beta$ (see section~\ref{sec:nutshell}).
      One should also keep in mind that the relative size of the largest
      cluster correlates with the density of the liquid phase (thus the
      order parameter) only on the critical isochore but is in general more
      readily interpreted as a measure of the mass fraction of the liquid phase.
\item {\bf Corrections to scaling:}
      The power laws listed in section~\ref{sec:nutshell} hold only
      asymptotically for $\epsilon \rightarrow 0$ while a description in a
      wider range is possible with correction to scaling terms~\cite{Wegner-72}
      or a crossover approach~\cite{Anisimov-92}.
      The size of the critical region depends strongly on the exact form of
      the interaction and is rather small in real liquids~\cite{Anisimov-92}
      but substantially larger in models with only next-neighbor or short
      range interactions.
      The leading correction term for the order parameter in the Ising
      model~\cite{Talapov-96} is not only substantially smaller in magnitude
      but has even a different sign as compared to typical
      liquids~\cite{Pestak-84}.
      It is therefore uncertain whether simple schematic models, like
      percolation or Ising, are adequate for the modeling of finite size
      and finite control parameter effects in the nuclear case.
\item {\bf Field gradients:}
      The hallmark of critical phenomena is scale invariance which at least
      requires homogeneous conditions throughout the system under study.
      A Coulomb or a radial flow~\cite{CRIS-Lacey} field will prevent the
      growth of fluctuations across the whole system much like the chemical
      potential gradient caused by the gravitational field is limiting the
      usable sample size in earthbound experiments on
      liquids~\cite{Moldover-79}.
\end{licompact}
All those points will have to be addressed before the experimental results can
be connected to the parameters of bulk nuclear matter in a quantitative way.
%
%
\section*{Acknowledgments}
J.P. and M.B. acknowledge the financial support of the Deutsche
Forschungsgemeinschaft under Contract Nos. Po256/2-1 and
Be1634/1-1, respectively.
This work was supported in part by the European Community under Contracts
ERBCHGE-CT92-0003 and ERBCIPD-CT94-0091.
We like to thank D. Stauffer for bringing Ref.~\cite{Moukarzel-89} to our
attention.
%
%


\begin{thebibliography}{99}
\bibitem{RNG-remark} This type of generator was shown to be
    adequate~\cite{Vattulainen-95} for the Wolff algorithm which is especially
    sensitive to correlations in random numbers~\cite{Ferrenberg-93}.
\end{thebibliography}
\end{document}